\documentclass{aa}
\usepackage{graphicx}
\begin{document}
\title{Far and mid Infrared Observations of Two Ultracompact H\,{\sc ii} 
Regions and One Compact CO Clump}
\author{R. P. Verma\inst{1} \and 
 S. K. Ghosh\inst{1}  \and  B. Mookerjea\inst{1,2}  \and T. N.
 Rengarajan\inst{1,3}}

\offprints{R.P. Verma, \email{vermarp@tifr.res.in}}

\institute{Tata Institute of Fundamental Research,
Homi Bhabha Road, Bombay 400 005, India \and
Now at I. Physikalisches Institut, Universitaet zu Koeln,
D 50937 Koeln, Germany \and
Now at Instituto National de Astrofisica, Optica, y Electronica, 
Puebla, Mexico}
\date{Accepted 15 November 2002}
\abstract{ Two ultracompact H\,{\sc ii} regions (IRAS 19181+1349 and
20178+4046) and one compact molecular clump (20286+4105) have been
observed at far infrared wavelengths using the TIFR 1 m balloon-borne
telescope and at mid infrared wavelengths using ISO. Far infrared
observations have been made simultaneously in two bands with effective
wavelengths of $ \sim150$ and $ \sim$210 $\mu$m, using liquid $^{3}$He
cooled bolometer arrays. ISO observations have been made in seven spectral
bands using the ISOCAM instrument; four of these bands cover the emission
from Polycyclic Aromatic Hydrocarbon (PAH) molecules. In addition, IRAS
survey data for these sources in the four IRAS bands have been processed
using the HIRES routine. In the high resolution mid infrared maps as well as
far infrared maps multiple embedded energy sources have been resolved.
There are structural similarities between the images in the mid infrared and
the large scale maps in the far infrared bands, despite very different
angular resolutions of the two. Dust temperature and optical depth
($\tau_{150}$) maps have also been generated using the data from
balloon-borne observations. Spectral energy distributions (SEDs) for these
sources have been constructed by combining the data from all these
observations. Radiation transfer calculations have been made to understand
these SEDs. Parameters for the dust envelopes in these sources have been
derived by fitting the observed SEDs.  In particular, it has been found
that radial density distribution for three sources is diffrent. Whereas in
the case of IRAS 20178+4046, a steep distribution of the form r$^{-2}$ is
favoured, for IRAS 20286+4105 it is r$^{-1}$ and for IRAS 19181+1349 it
the uniform distribution (r$^{0}$).  Line ratios for PAH bands have
generally been found to be similar to those for other compact H\,{\sc ii}
regions but different from general H\,{\sc ii} regions.
}

\titlerunning{Far and mid infrared observations ...}
\authorrunning{R.P. Verma et al.}
\maketitle

\section{Introduction}

Ultracompact H\,{\sc ii} regions are a class of interesting astronomical
sources that represent early stages of deeply embedded high mass (O or
early B) stars.  Therefore their study can provide vital information about
high mass star formation as well as their interaction with the parent
molecular cloud. They are also often associated with OH/H$_2$O
masers. Being deeply embedded in dust, almost all of their energy is
emitted in the infrared waveband. In this paper we report far and mid
infrared observations of two ultracompact H\,{\sc ii} regions, IRAS
19181+1349 and IRAS 20178+4046. We also report observations for a compact
molecular clump, IRAS 20286+4105, which is similar to the above sources in
many respects.

IRAS 19181+1349 and IRAS 20178+4046 are ultracompact H\,{\sc ii} regions
(Kurtz et al. 1994) with rising far infrared spectra between 60 and 100
$\mu$m. Radio observations with the VLA exhibit multiple compact sources in
the field of IRAS 19181+1349 (Kurtz et al. 1994; Zoonematkermani et al. 1990).
Extended radio emission at 3.6 cm has also been seen in both the sources;
however only for IRAS 19181+1349 is the extended emission likely to be
connected with the ultracomact H\,{\sc ii} region (Kurtz et al. 1999). 
Molecular lines of CO and CS have been observed towards both of these
sources (Shepherd and Churchwell, 1996; Bronfman et al. 1996).
IRAS 19181+1349 is also associated with H$_{2}$O and OH masers (Forster
and Caswell 1989, Palagi et al. 1993); formaldehyde and methanol masers
have not been detected (Caswell et al. 1995; Mehringer et al. 
1995). No maser has been found to be associated with IRAS 20178+4046. 
Between 8 and 22 $\mu$m IRAS low resolution spectrometer (LRS) spectrum is 
available for IRAS  20178+4046 (Volk and Cohen 1989) and groundbased
intermediate resolution spectroscopic observations have been made between
3 to 13 $\mu$m (Faison et al. 1998). In the latter, PAH
features\footnote{These features are also called unidentified
infrared bands (UIBs) meaning that the PAH origin is not completely settled.
However we shall consistently use the term PAH.} at 3.3
$\mu$m, 8.7 $\mu$m and 11.3 $\mu$m, [Ne II] line at 12.8 $\mu$m and
silicate feature at 9.7 $\mu$m have been detected.  Based on radio
recombination observations, the kinematic distance of IRAS 19181+1349 has
been variously determined as 9.7 kpc to 11.8 kpc (Caswell et al. 1975,
Kurtz et al. 1994). For IRAS 20178+4046 the kinematic distances from 1.5 kpc
to 3.3 kpc have been used (Kurtz et al. 1994, Casoli et al. 1986). We use
a distance of 9.7 kpc for IRAS 19181+1349 and 3.3 kpc for IRAS 20178+4046. 

IRAS 20286+4105 is an unusual compact CO source (Odenwald 1989) with
infrared colours similar to ultracompact H\,{\sc ii} regions (according to
the criteria of Wood and Churchwell, 1989), associated with a young stellar
object (YSO). It is located at the centre of a thin luminous ring of
$\sim 1\farcm 5$ diameter in the visible (Odenwald and Schwartz 1989). Other
molecules like NH$_3$, H$_2$O and CS have also been observed towards this
source (Molinari et al. 1996, Bronfman et al. 1996). Near infrared
imaging of this source in J, H and K bands has been done by Comeron and
Torra (2001). LRS spectrum of IRAS 20286+4105 (Olnon et al. 1986) shows a
broad absorption feature at 11.3 $\mu$m. It has also been associated with
H$_2$O masers (Codella et al. 1996). The kinematic distance for this source
is 3.7 kpc (Molinari et al. 1996).

As part of a programme to obtain far infrared images of star forming
regions (Ghosh et al. 1988, 1989a, 1989b, 1990, 2000; Mookerjea et al. 1999)
we have mapped these three regions simultaneously in two far infrared bands
using the 1 m balloon-borne telescope of the Tata Institute of Fundamental
Research (TIFR). The effective wavelengths of the two bands are $\sim
150~\mu \rm m$ and $\sim 210~\mu \rm m$ and the angular resolution acheived
is $\sim 1 \arcmin$ at both wavelengths. We have also procured HIRES
(Aumann et al. 1990) processed IRAS maps for these sources in all the four
IRAS bands. To understand the role of transient heating of PAH molecules,
we have imaged the central $3 \arcmin \times 3 \arcmin$ regions of these
sources in seven mid-infrared bands using ISOCAM in the ISO\footnote{Based
on observations with ISO, an ESA project with instruments funded by ESA
Member States (especially the PI countries: France, Germany, the Netherlands
and the United Kingdom) with the participation of ISAS and NASA.} satellite.

The observational details are given in Sect. 2. The results are presented
in Sect. 3. Spectral energy distributions (SEDs) have been constructed by
combining our results with other observations. To understand the infrared
emission from these sources, radiation transfer calculations have been
performed. The schemes of radiation transfer calculations are presented in
Sect. 4. Discussion of sources is presented in Sect. 5. Conclusions are
summarized in Sect. 6. Preliminary results of these observations have been
given in Verma et al. (1999a,b) and Karnik (2000).

\section{Observations}
\subsection{Balloon-borne observations}
Balloon-borne far infrared observations were made with the TIFR 1 m
telescope during a balloon flight on February 20, 1994 from
Hyderabad (lat. $17\fdg47$ N, long. $78\fdg57$ E), India. The details of the 
telescope and the observational procedures have been described by 
Ghosh et al. (1988); here only some basic features will be mentioned. The 
telescope is a 100 cm aperture f/8 Cassegrain system. It can be pointed towards 
a desired direction in the sky with an rms pointing stability of better than 
$\sim 0\farcm3$. The incident radiation is chopped by vibrating the secondary 
mirror along the cross elevation direction at a frequency of 10 Hz and a 
throw of $4\farcm2$.  The detectors consist of two arrays (each 2$\times$3) 
of composite Si bolometers cooled to 0.3 K by a closed cycle
$^3$He refrigerator. The field of view for each detector is $1\farcm6$.
The observations were made simultaneously in two bands with effective
wavelengths of $\sim 150 \, \mu$m and 210\,$\mu$m. The two bands are
separated by the use of cool CsI restrahlen beam-splitter which reflects
the radiation with wavelengths between 85\,$\mu$m and 180\,$\mu$m and
transmits the radiation with wavelengths shorter than 85\,$\mu$m and
longer than 180\,$\mu$m. The radiation with wavelengths shorter than
85\,$\mu$m is blocked by other blocking filters. Some of the details of the
photometer are given in Verma et al. (1993). Measured transmissions for
both the bands as well as the effective wavelengths and the relative
responsivities (which depend on the source spectrum) are given in
Ghosh et al. (2000). For a dust temperature of 30 K and a gray body
spectrum with $\lambda ^{-2}$ emissivity, the effective wavelenghts for
the two bands are 148 $\mu$m and 209 $\mu$m. For simplicity, we convert
the flux densities in the two bands to values at wavelengths of 150 and
210 $\mu$m respectively.

Mapping was done in a raster mode with a scan speed of $0\farcm8~\rm s^{-1}$
along the cross elevation axis. The boresight of the
telescope was determined for each signal sample using the data from the
star tracker and the rate gyroscopes. For each band, signals from all six
detectors were used to form a two dimensional signal grid in instrument
coordinates (elevation, cross-elevation) with a cell size of $0\farcm3
\times 0\farcm3$.  Data have been deconvolved using a Maximum Entropy
Method (MEM) similar to that of Gull and Daniell (1978). Absolute positions
were calibrated using observations of catalogued stars with an optical
photometer in the focal plane; field of which is offset with respect to the
infrared field. The calibrated absolute positions are accurate to 
1$\arcmin$. Jupiter was observed at the begining of the flight as
well as towards the end of the flight; Jupiter observations were used for
flux calibration. As the Jupiter size ($0\farcm6$) was much less than the
field of view ($1\farcm6$), Jupiter observations were also used for the
determination of the point spread function (PSF) for MEM deconvolution. For
the purpose of flux calibration we have assumed Jupiter to be a black body
with a temperature of 127 K. The final resolution in the deconvolved maps as
given by the FWHM (El. $\times$ XEl.) of Jupiter is $ 1 \farcm 0 \times 1
\farcm 3$ in both the bands.

    Taking advantage of the simutaneous observations in the two bands, 
with almost identical field of view, we have generated maps of the dust 
temperature (T$_{d}$) and optical depth at 150 $\mu$m ($\tau_{150}$). The
procedure for the generation of these maps is described in Mookerjea et al.
(2000). For these maps we have assumed the dust emissitivity of
$\epsilon_{\lambda}~ \propto \lambda^{-2}$ and a binning of 3 pixel by 3
pixel has been effected for both the bands before computing the T$_{d}$
and $\tau_{150}$. The temperature and optical depth maps have been
restricted to regions where the flux densities in both the maps are well
within the dynamic range.

\subsection{IRAS observations}
IRAS survey data for these three sources were processed using the HIRES
routine at the Infrared Processing and Analysis Center (IPAC), Caltech
\footnote{ IPAC is funded by NASA as part of the IRAS extended mission 
program under contract to JPL} 
to obtain high resolution images in all the four IRAS bands. Details of
the HIRES routine have been given in Aumann et al. (1990). Angular
resoluions in different images along observed source sizes are given in Table
1. The HIRES maps have been used to obtain flux densities and angular sizes
in the IRAS bands.

\begin{table*}[tb]
  \caption{ Map resolutions and source sizes}
  \begin{center}
    \leavevmode
    {\footnotesize
\begin{tabular}{llcccccc} \hline
Source &   & \multicolumn{6} {c}{FWHM Sizes in arcminutes (minor
$\times$ major)}\\
\cline{3-8}
 & & \multicolumn{4} {c} {HIRES-IRAS} & \multicolumn{2} {c} {TIFR} \\
\cline{3-8}
   &   & 12\,$\mu$m & 25\,$\mu$m & 60\,$\mu$m & 100\,$\mu$m & 
          150\,$\mu$m &   210\,$\mu$m\\ \hline
19181+1349 & Map resolution & 0.5$\times$1.2 & 0.5$\times$1.0 &
1.1$\times$1.6 & 2.0$\times$2.3 & 1.0$\times$1.3 & 1.0$\times$1.3 \\
           & Source size & 1.3$\times$2.8 & 0.9$\times$2.1 &
1.3$\times$2.1 & 2.3$\times$3.1 & 2.1$\times$3.7 & 2.2$\times$2.3 \\
  & & & & & & &  \\
20178+4046 & Map resolution & 0.5$\times$1.1 & 0.6$\times$1.2 &
1.1$\times$1.8 & 1.9$\times$2.4 & 1.0$\times$1.3 & 1.0$\times$1.3 \\
           & Source size & 0.6$\times$0.6 & 0.5$\times$0.7 &
1.0$\times$2.0 & 1.9$\times$2.2 & 1.1$\times$1.5 & 1.5$\times$1.5 \\
           & Model size & 0.6$\times$1.3 & 0.6$\times$0.6 &
1.1$\times$2.1 & 1.9$\times$2.5 & 1.2$\times$1.4 & 1.2$\times$1.5 \\
 & & & & & & &  \\
20286+4105 & Map resolution & 0.5$\times$1.0 & 0.6$\times$1.2  &
1.0$\times$1.1 & 1.9$\times$2.4 & 1.0$\times$1.3 & 1.0$\times$1.3 \\
           & Source size & 0.8$\times$1.3 & 0.6$\times$0.9 &
1.1$\times$1.5 & 1.8$\times$2.0 & 0.9$\times$1.2 & 1.2$\times$1.5 \\
\hline
      \end{tabular}
}  
\newline
\end{center}
\end{table*}

\subsection{ISO observations}
  The central portions of the sources were observed at mid infrared
wavelengths using ISOCAM. An area of $3 \arcmin\times 3 \arcmin$ around
each source has been imaged in seven spectral filters. The filters are SW2,
(with central wavelengths of 3.30 $\mu$m; bandwidths of 0.20 $\mu$m), 
SW6 (3.72; 0.55), LW4 (6.00; 1.0), LW5 (6.75; 0.5), LW6 (7.75; 1.5), LW7
(9.62; 2.2) and LW8 (11.4; 1.3) (ISOCAM Observer's Manual, 1994). Out
of these filters, four (SW2, LW4, LW6 and LW8) include emission from
PAH bands whereas the remaining three bands are used as comparison
for continuum emission. The
pixel size used for imaging was $6\arcsec \times 6 \arcsec$.  Integration
time per exposure was 2.0 s for SW filters and 0.28 s for LW filters.
A sufficient number of exposures were taken for detector stabilization
before the actual observations. The data were analyzed using Cam
Interactive Analysis (CIA) version 3.0.

\section{Results}
\subsection{IRAS 19181+1349}
Around IRAS 19181+1349 an area of $\sim 24 \arcmin \times 30 \arcmin$
was scanned twice and the resulting deconvolved maps in the two far infrared
bands are shown in Fig. 1. For this region no suitable star was seen by the
optical photometer, therefore the expected error in absolute position is
larger ($\sim2\arcmin$). We have made corrections to the coordinates to
match the position of the main peak with that in the IRAS HIRES map at
100 $\mu$m. As can be seen from the figure, there are multiple sources in
both the maps. Whereas in the longer wavelength band the source is resolved
into three components and a lobe in the west, in the shorter wavelength
band it is resolved into two components. The deconvolved FWHM size of
IRAS 19181+1349 at 150 $\mu$m is $2 \farcm 1 \times 3 \farcm 7$ and at
210 $\mu$m it is $2 \farcm 2 \times 2 \farcm 3$ ; the latter encloses the
three components. The second component seen at 150 $\mu$m is not covered
by ISOCAM observations. Besides IRAS 19181+1349, many other sources are
seen in the maps. Among these are sources corresponding to IRAS 19175+1357
and IRAS 19175+1342. Also, there is a lot of diffuse emission in both
the bands. Integrated flux densities around the peaks are given in Table 2.
%
\begin{figure*}
\centering
\includegraphics[width=16.0cm]{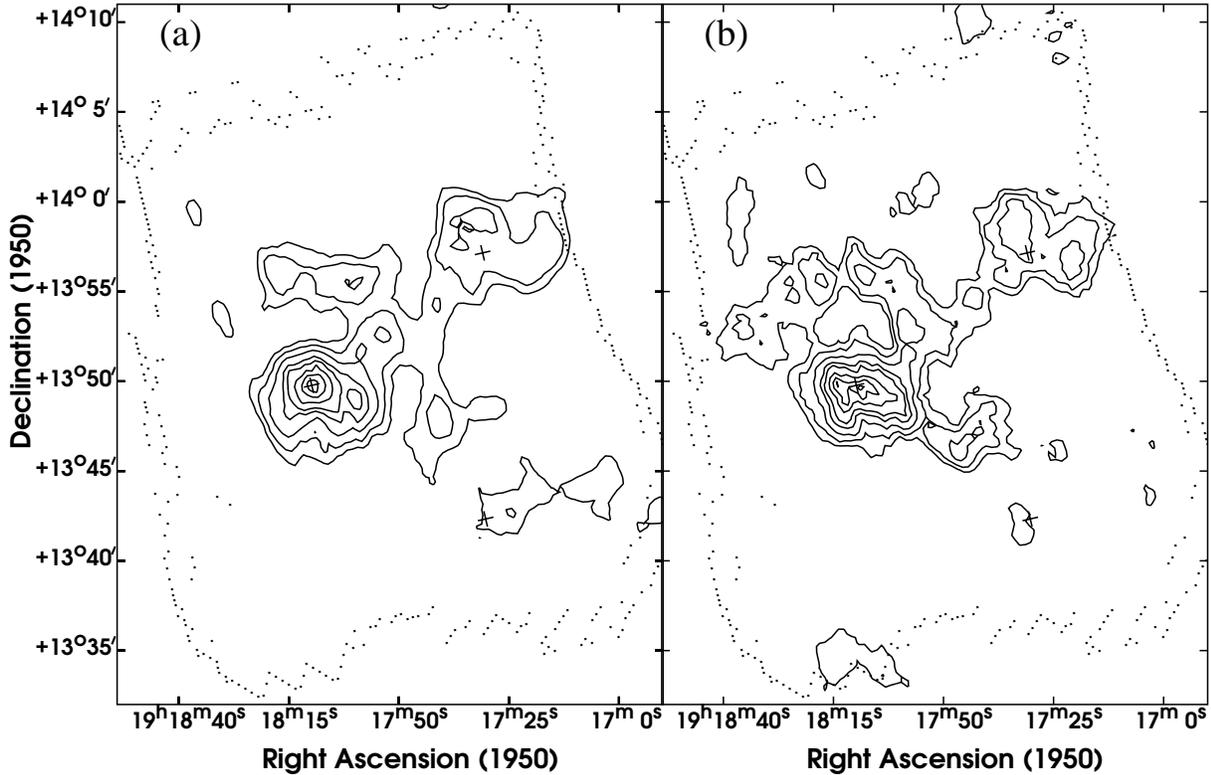}
\caption{Intensity maps of the region around IRAS 19181+1349 at {\bf (a)} 
150 $\mu$m and {\bf (b)} 210 $\mu$m. The dotted boundary marks the area
scanned. Crosses mark the positions of IRAS PSC sources. The contour
levels are at  0.90,  0.70, 0.50,
 0.30, 0.20, 0.10, 0.05 and 0.025 of the peak intensity. The peak
intensities are 723 Jy arcmin$^{-2}$ at 150 $\mu$m and 390 Jy
arcmin$^{-2}$ at 210 $\mu$m.
}
\end{figure*}

\begin{table*}[tb]
  \caption{ Far infrared flux densities of the sources}
  \begin{center}
    \leavevmode
    {\footnotesize
\begin{tabular}{lcccccccc} \hline
Source &  RA & Dec. & \multicolumn{6} {c}{Flux Density in Jy$^{*}$}\\
\cline{4-9}
 & & & \multicolumn{4} {c} {HIRES-IRAS} & \multicolumn{2} {c} {TIFR} \\
\cline{4-9}
   & (1950)  & (1950)  & 12\,$\mu$m & 25\,$\mu$m & 60\,$\mu$m & 100\,$\mu$m & 
          150\,$\mu$m &   210\,$\mu$m\\ \hline
19181+1349 & 19 18 09.6 & 13 49 46 & 156 & 669 & 6443 & 7770 & 3939 & 2590 \\  
19181+1349E& 19 18 12.7 & 13 49 55 & 46 & 267 &  $<$2427  &
 $<$2236  &  $<$1096  & 656  \\
19181+1349W& 19 18 07.4 & 13 49 39 & 49 & 265 &  $<$1796
&   $<$1679  &  $<$1496  & 827  \\
19176+1359 & 19 17 36.8 & 13 59 16 & 14  & 47 &  $<$473 &  $<$846
  &   572  &  351  \\
19175+1357 & 19 17 32.7 & 13 57 16 & 22 & 95 &   $<$877  & 1254 &
 $<$680   &  $<$367  \\
 & & & & & & & &  \\
20286+4105 & 20 28 38.5  & 41 05 43  & 39 & 170 & 1071 & 1348 & 1249 & 902 \\
 & & & & & & & &  \\
20178+4046 & 20 17 48.6 & 40 46 44  & 114 & 748 & 2932 & 3325 & 2244 & 1199\\
\hline
      \end{tabular}
}  
\newline
\end{center}
  $^*${\footnotesize The flux densities have been integrated 
in a circle of 4$\arcmin$ diameter except for IRAS 19181+1349 where
integration has been done in a circle of 5$\arcmin$ diameter; for the two
components 19181+1349E and 19181+1349W the integrations have been done in
a circle of 2$\arcmin$ diameter.}\\

\end{table*}
 The dust temperature (T$_{d}$) and optical depth ($\tau_{150}$) maps
for IRAS 19181+1349 are shown in Fig. 2. It is seen that these maps are
complex. There are multiple peaks in temperature which are shifted with
respect to the main intensity peak by  $1\arcmin$ to $4\arcmin$. Optical
depth map has three peaks.

\begin{figure*}
\centering
\includegraphics[width=16.0cm]{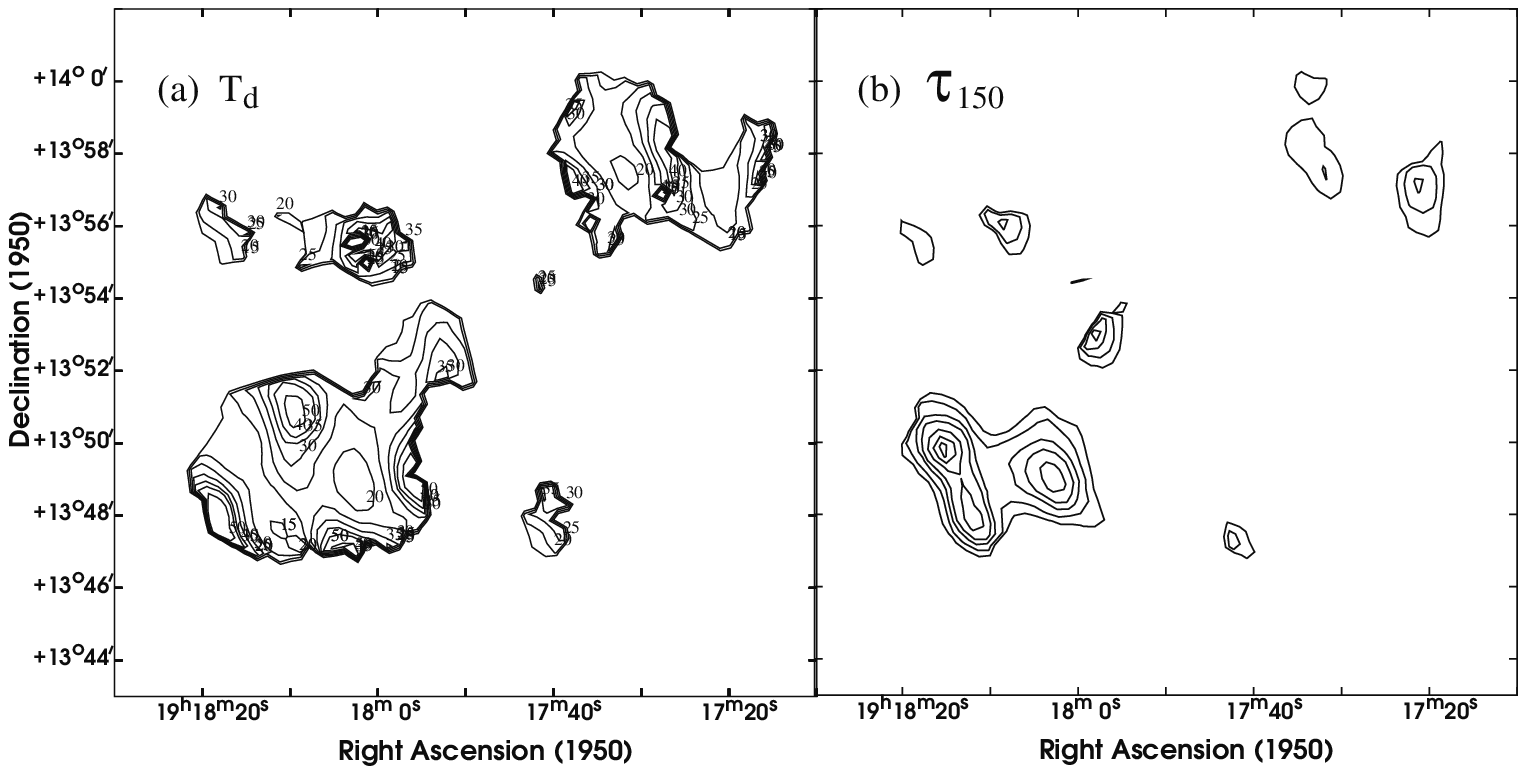}

\caption{Dust temperature (T$_{d}$) and optical depth
($\tau_{150}$) distribution for the region around IRAS 19181+1349. {\bf (a)}
Temperature distribution. The contours are at the levels of 
50, 40, 35, 30, 25, 20 and 15 K. {\bf (b)} Optical depth distribution. The
contours are at 90, 70, 50, 30, 20, 10 and 5\% of the peak
value which is 0.1.}
   
\end{figure*}

    The HIRES processed maps of IRAS 19181+1349 in the four IRAS bands,
derived from the survey data are shown in Fig. 3. In HIRES maps also the
source is resolved at all four wavelengths. It shows two components at
12 and 25 $\mu$m but at 60 and 100 $\mu$m only one component is seen.
However, as seen from Table 1, even at 60 and 100 $\mu$m the source is
extended with FWHM of $1\farcm3\times2\farcm1$ and $2\farcm3\times3\farcm1$
respectively as compared to the resolution of $1\farcm1\times1\farcm6$ and
$2\farcm0\times2\farcm3$.

\begin{figure*}
\centering
\includegraphics[width=16.0cm]{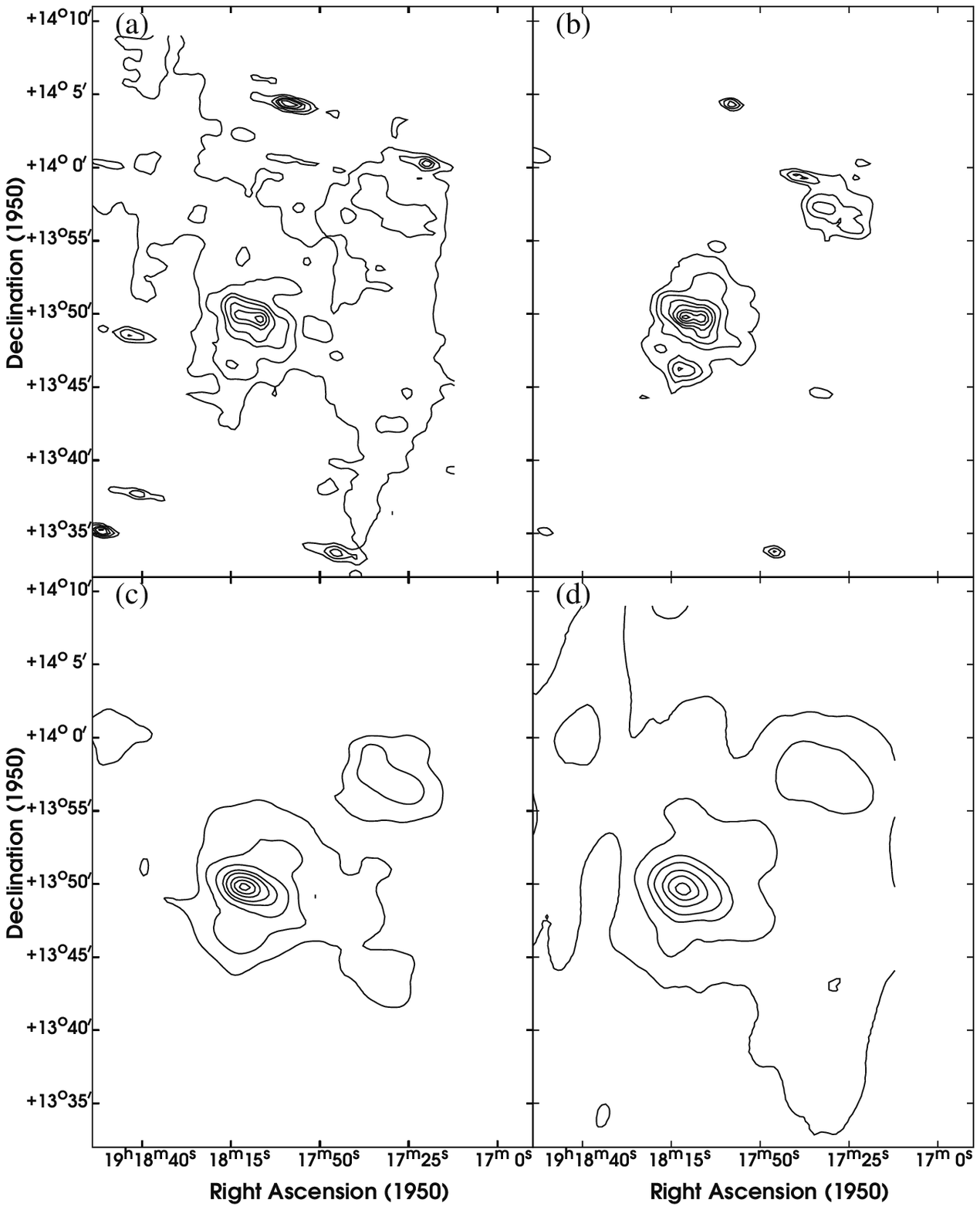}

   \caption{The HIRES processed intensity maps of the region around IRAS
19181+1349 at {\bf (a)} 12 $\mu$m, {\bf (b)} 25 $\mu$m, {\bf (c)} 60 $\mu$m and
{\bf (d)} 100 $\mu$m. The contour levels are at  0.90,  0.70,  0.50, 0.30,
0.20, 0.10, and 0.05 of the peak intensity for 12 $\mu$m; for other
wavelengths, besides these there are additional contours at 0.025 of the peak.
The peak intensities are 26.4, 181, 1580 and 1090 Jy arcmin$^{-2}$ at 12,
25, 60 and 100 $\mu$m respectively.}

\end{figure*}

    Maps generated from ISOCAM images in the seven bands are shown in
Fig. 4. The images in the SW filters have been physically shifted such
that the positions of the main peaks in these images coincide with the
respective positions in the LW filters. This was done to correct for
the lens wheel jitter. This procedure has also been followed for the SW
images of the other two sources. The required shift is $\sim 20 \arcsec$.
It can be seen from this figure that IRAS 19181+1349 consists of multiple
components in all the filters. However the secondary peaks in the two SW
filters do not appear in the LW filters and therefore do not seem to be
associated with the IRAS source. The components in the LW filters can be
divided into two groups separated by $\sim 1\farcm5$, mainly in the E--W
direction. This feature is also seen at 12 and 25 $\mu$m in the HIRES maps
and at 210 $\mu$m in the TIFR map. Integrated flux densities of strong
components are given in Table 3.

\begin{table*}[tb]
  \caption{ Flux densities of sources in the ISOCAM images}
  \begin{center}
    \leavevmode
  {\footnotesize
\begin{tabular}{cccccccccc} \hline
IRAS & RA & Dec. & \multicolumn{7} {c}{Flux Density in Jy$^{*}$}\\
\cline{4-10}
Source  & (1950)  & (1950) & SW2 & SW6 & LW4 & LW5 & LW6 & LW7 & LW8\\ 
        & & & 3.30\,$\mu$m & 3.72\,$\mu$m &  6.00\,$\mu$m&  6.75\,$\mu$m & 
           7.75\,$\mu$m   &  9.62\,$\mu$m &  11.4\,$\mu$m\\ 
     & & & PAH &     & PAH &     & PAH &     & PAH\\ \hline
19181+1349  & 19 18 05.4  &+13 49 22 & -- & -- & 0.81
 & 1.34   &   2.6  & -- & 1.29 \\
            & 19 18 05.8  &+13 50 32 & 0.10${^\#}$  & --  & -- & --   &
  --  & --  & --  \\ 
           & 19 18 06.8  &+13 49 40 & -- & -- & 0.73
     & 1.22   &   2.4 & 1.29 & 3.0 \\
           & 19 18 07.6  &+13 48 54 & 0.17${^\#}$ & 0.19${^\#}$ & -- & --
  &  --  & --  & --  \\ 
            & 19 18 11.2  &+13 49 33 & 0.19 &  --    & 1.21 & 1.70  &
 3.6  &  --  &  1.5 \\ 
            & 19 18 12.2  &+13 50 52 & 0.09 & -- & -- & --   &   --  &
--  & --  \\ 
            & 19 18 12.5  &+13 51 24 & 0.21 & 0.20 &  -- & --   &   --  &
--  & --  \\ 
            & 19 18 12.6 & +13 50 10 & 0.77 & 0.56 & 2.5 & 2.8 & 5.9 &
1.90 & 3.4\\ 
           &19 18 13.0  &+13 49 59 & 0.23${^\#}$ & 0.13${^\#}$ &
   1.27${^\#}$   & 1.51${^\#}$   &   2.85${^\#}$  & -- & -- \\
            & 19 18 13.5 & +13 49 24 & -- & -- & 0.78 & 0.96 & 2.0 &
-- & --\\ 
  &  &  & & & & & & & \\ 
20178+4046 & 20 17 53.3 & +40 47 03 &  1.8 & 0.70 & 11.6 & 10.7 & 30.4 &
11.0 & 20.9\\       
          & 20 17 55.0 & +40 47 01 &  0.30${^\#}$ & -- & 2.7${^\#}$ &
  2.7${^\#}$ & 7.5${^\#}$ & 2.7${^\#}$ & 4.6${^\#}$\\       
  &  &  & & & & & & & \\ 
20286+4105 & 20 28 40.3 & +41 05 14 & 0.71 & 0.23 & 4.4 & 4.8 & 11.5 &
3.5 & 6.3\\ 
           & 20 28 40.3 & +41 05 45 & 0.25 & 0.13 & 3.1 & 3.3 & 7.8 &
1.90 & 2.8\\ 
           & 20 28 42.5 & +41 05 44 & 0.20 & 0.12 & 2.9 & 2.4 & 3.5 &
1.21 & 1.50\\ 
 \hline
      \end{tabular}
}
\newline
  \end{center}
  $^*${\footnotesize The flux densities have been integrated in a circle
of $30\arcsec$ diameter around the peak except for those marked with $^\#$
where the integration is done over an area of $18\arcsec \times 18
\arcsec$ around the peak.}\\

\end{table*}

\begin{figure*}
\centering
\includegraphics[width=16.0cm]{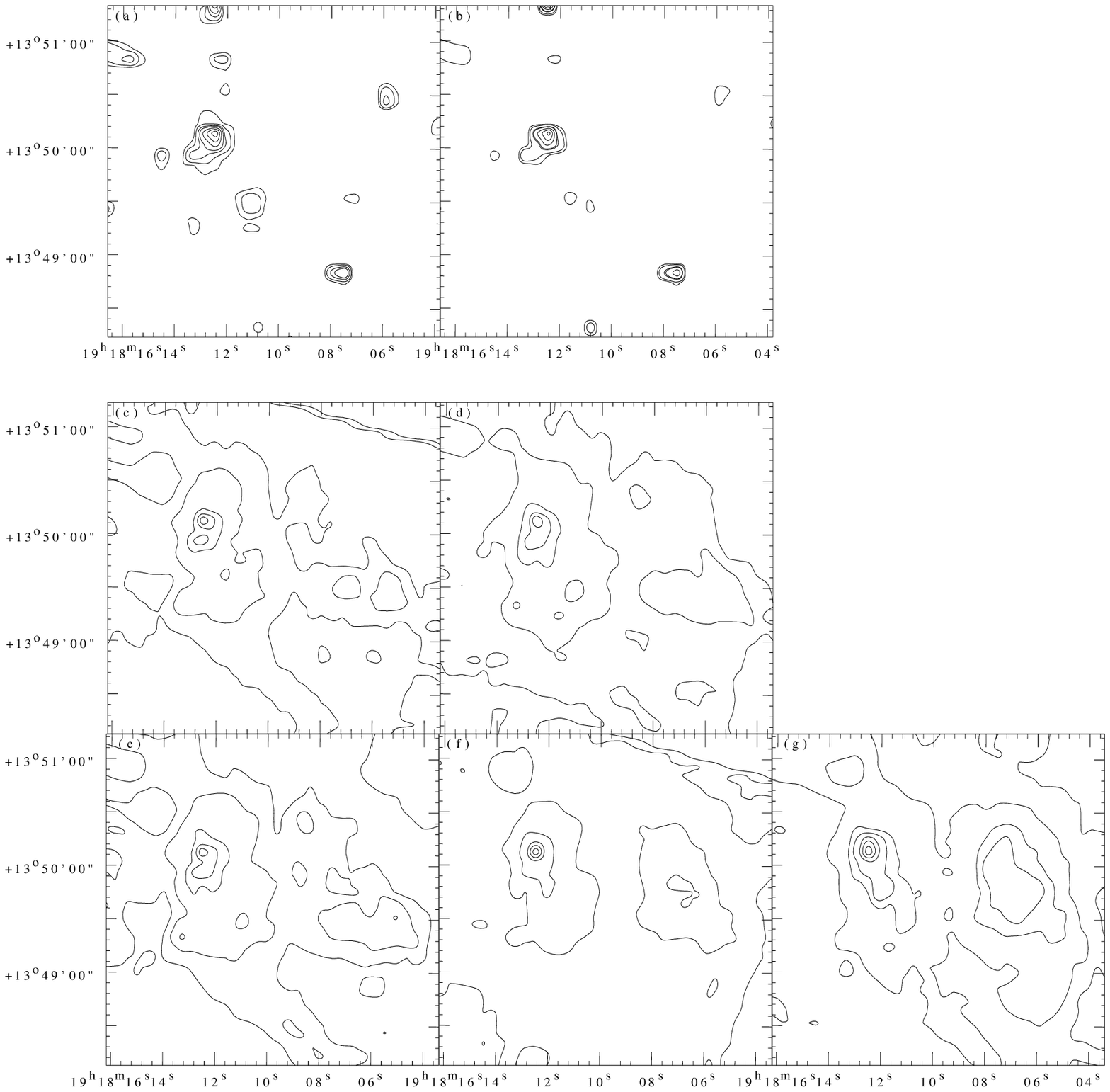}

   \caption{The maps of ISOCAM observations  of the region around IRAS
19181+1349 at {\bf (a)} 3.30 $\mu$m, {\bf (b)} 3.72 $\mu$m, {\bf (c)} 6.00
$\mu$m, {\bf (d)} 6.75 $\mu$m, {\bf (e)} 7.75 $\mu$m, {\bf (f)} 9.62 $\mu$m,
 and {\bf (g)} 11.4 $\mu$m.  The
contour levels are at 0.90, 0.70, 0.50, 0.30, 0.20, 0.10 and 0.05 of the
peak intensity. The peak intensities are 19.5, 17.8, 36.5, 37.7, 70.2,
37.8 and 58.7 Jy arcmin$^{-2}$  at  3.30, 3.72, 6.00, 6.75, 7.75, 9.62 and
11.4 $\mu$m respectively.}

\end{figure*}

\subsection{IRAS 20178+4046}
Around IRAS 20178+4046 an area of $\sim 30 \arcmin \times 30 \arcmin$
was scanned twice and the resulting deconvolved maps in the two far infrared
bands are shown in Fig. 5. This source does not show multiple components
in any of the two maps. It can be seen from Table 1 that the source is
partly resolved.
   
\begin{figure*}
\centering
\includegraphics[width=16.0cm]{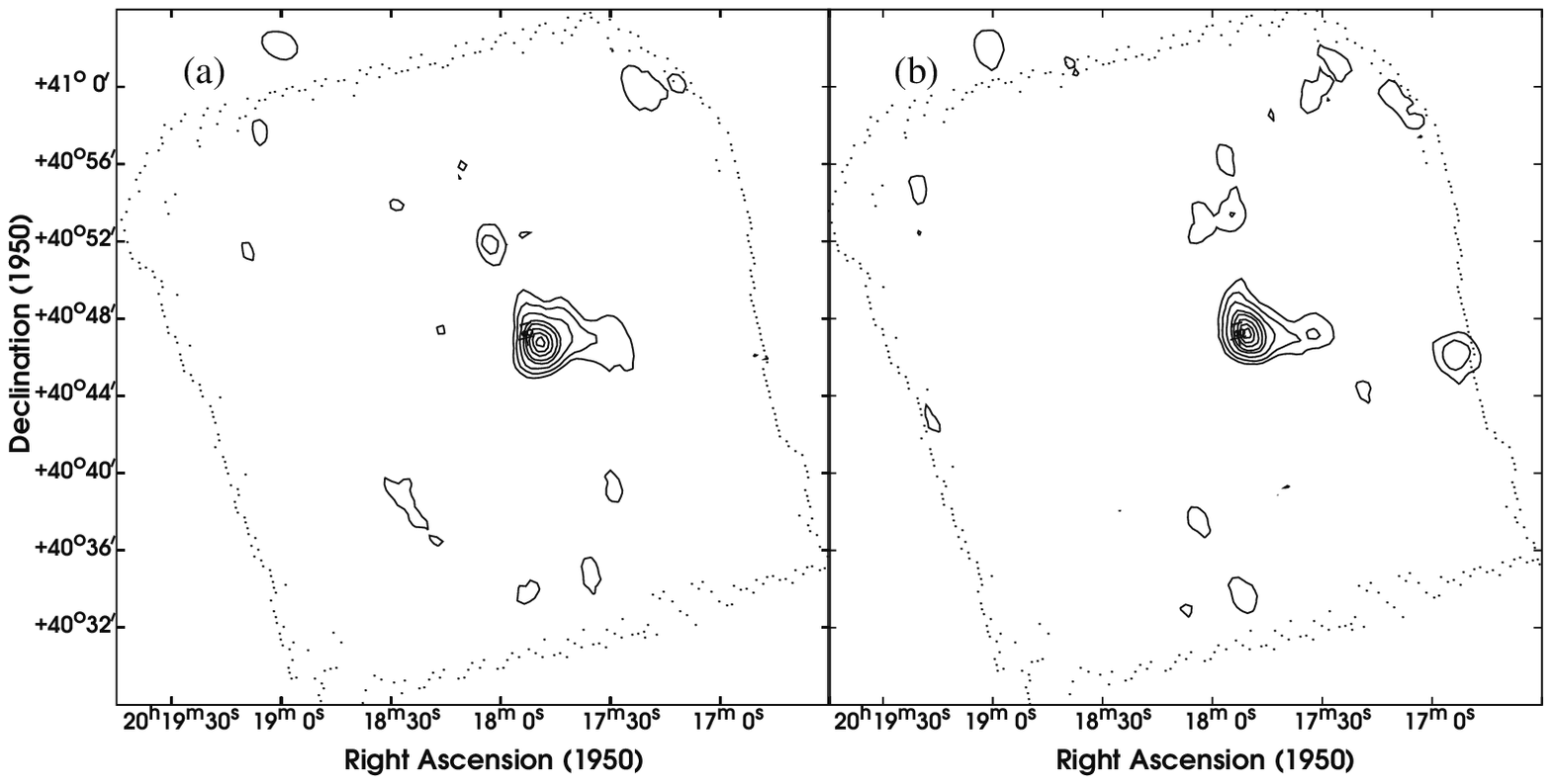}

   \caption{Same as figure 1 but for the region around IRAS 20178+4046.
The peak intensities are 978 Jy arcmin$^{-2}$ at 150 $\mu$m and 589 Jy
arcmin$^{-2}$ at 210 $\mu$m.}

\end{figure*}

 The dust temperature (T$_{d}$) and optical depth ($\tau_{150}$)
maps for IRAS 20178+4046 are shown in Fig. 6. It is seen that the peak
in optical depth occurs close to the peak in the intensity distributions.
The temperature distribution shows two peaks separated by $2 \farcm 4$.
 The total range in temperature is from 72 K to 21 K.

\begin{figure*}
\centering
\includegraphics[width=16.0cm]{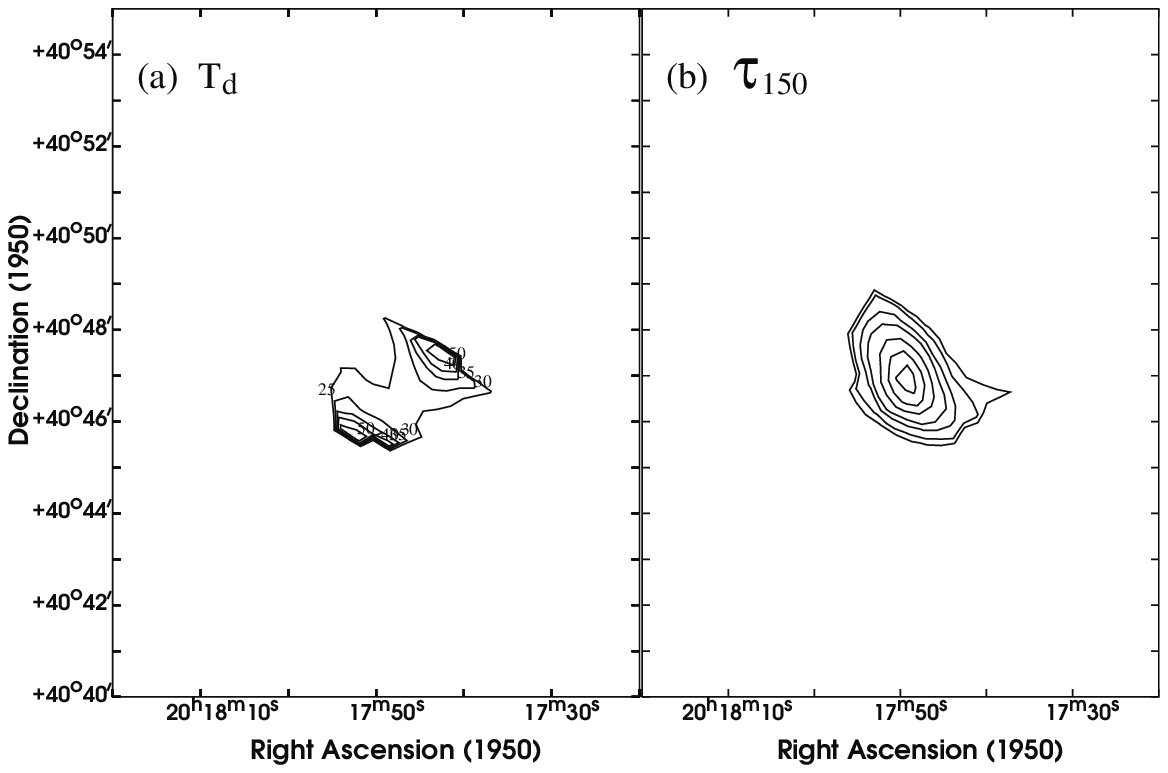}

\caption{Same as figure 2 but for the region around IRAS 20178+4046.
The temperature contours are at the levels of 50, 40, 35, 30 and 25 K
and for the optical depth distribution the peak value is 0.038.}

\end{figure*}

    The HIRES processed maps of IRAS 20178+4046 in the four IRAS bands,
derived from the survey data, are shown in Fig. 7. The central part of the
source is not resolved in any of the IRAS bands, as seen from Table 1. However
there is lower level extended emission in all the bands except at 25 $\mu$m.
   
\begin{figure*}
\centering
\includegraphics[width=16.0cm]{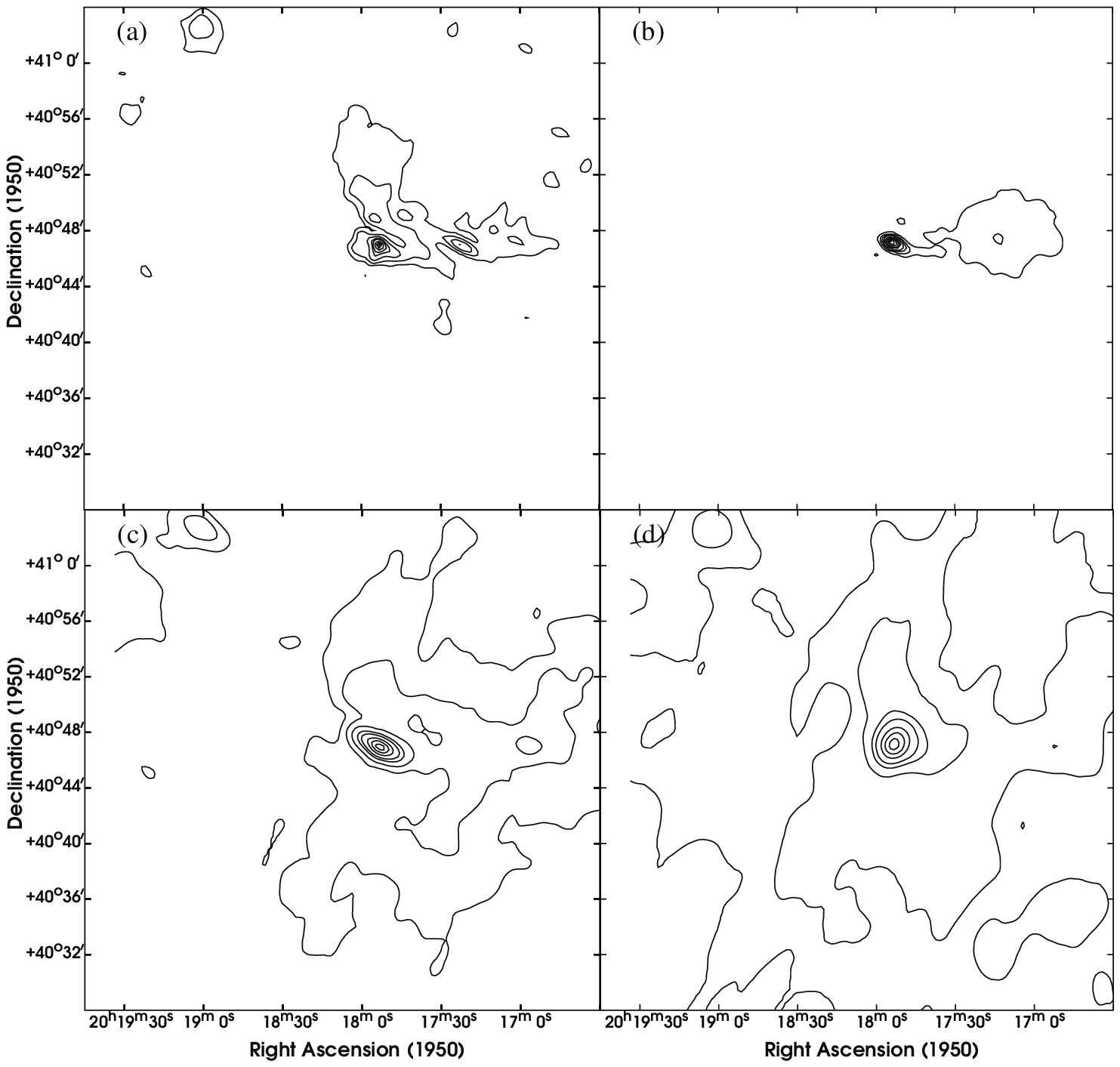}

   \caption{The HIRES processed intensity maps of the region around IRAS
20178+4046 at {\bf (a)} 12 $\mu$m, {\bf (b)} 25 $\mu$m, {\bf (c)} 60 $\mu$m
and {\bf (d)} 100 $\mu$m. The contour levels are at  0.90, 0.70, 0.50, 0.30,
0.20, 0.10, 0.05
and 0.025 of the peak intensity; for 25 $\mu$m additional contour at 0.01
of the peak has been shown. The peak intensities are 89.7, 1490, 1110 and
737 Jy arcmin$^{-2}$ at 12, 25, 60 and 100 $\mu$m respectively.}

\end{figure*}

    Maps generated from the ISOCAM images in the seven bands are shown in
Fig. 8. It can be seen from this figure that IRAS 20178+4046 consists of
only one component in all the filters. The contours in all the LW filters
are remarkably similar to each other, showing a lobe towards the south.

\begin{figure*}
\centering
\includegraphics[width=17.0cm]{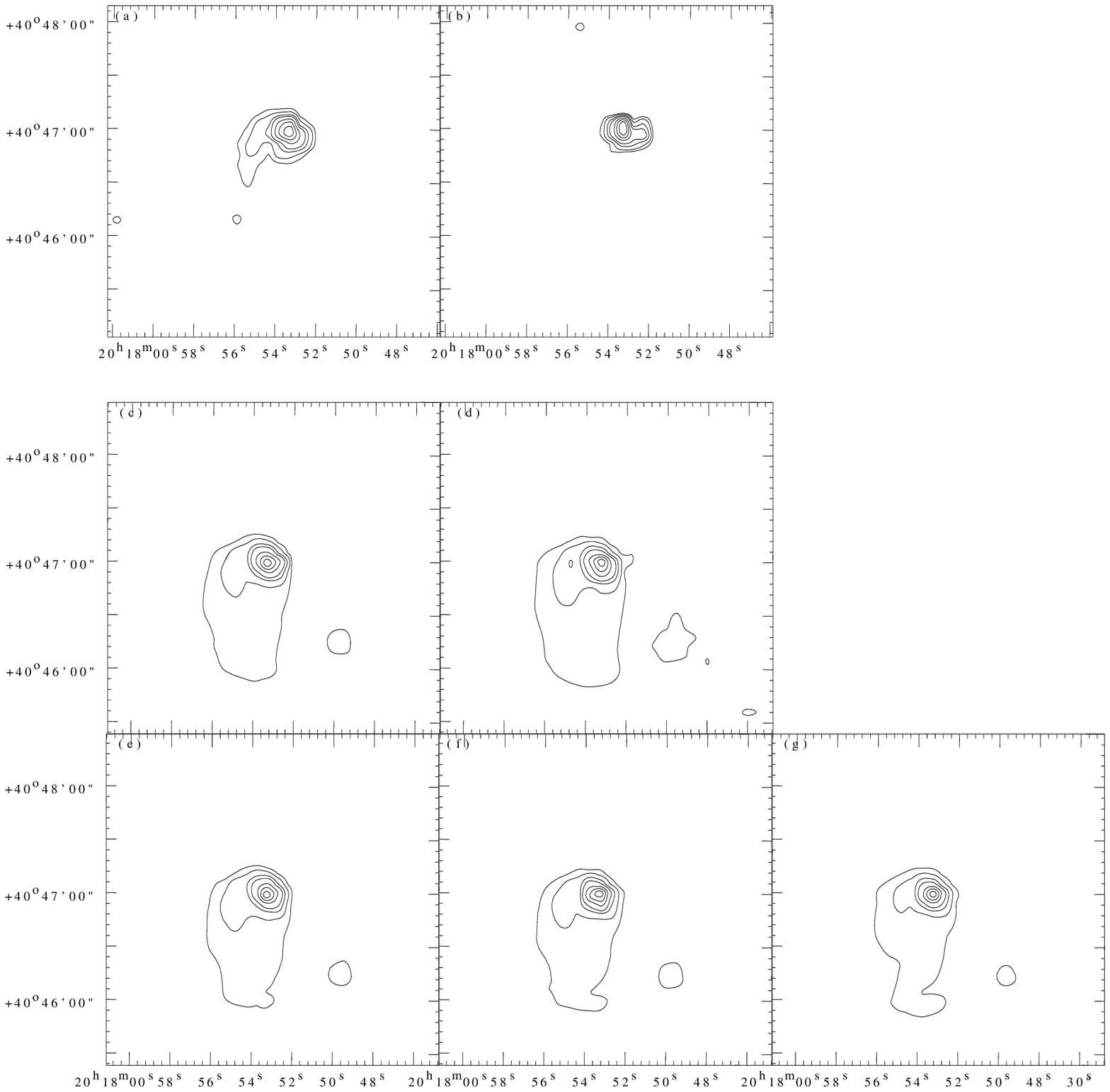}

\caption{Same as figure 4 but for the region around IRAS 20178+4046. 
The peak intensities are 30.3, 15.2, 232, 214, 617, 239 and 505 Jy
arcmin$^{-2}$ at 3.30, 3.72, 6.00, 6.75, 7.75, 9.62 and 11.4 $\mu$m
respectively.}

\end{figure*}

\subsection{IRAS 20286+4105}
Around IRAS 20286+4105  an area of $\sim 16 \arcmin \times 30 \arcmin$
was scanned twice and the resulting deconvolved maps in the two far infrared
bands are shown in Fig. 9. As in the case of IRAS 20178+4046, this source
also has only one component in both the bands. The source is resolved at
210 $\mu$m but at 150 $\mu$m it is unresolved. At 210 $\mu$m it shows an
extension in the same direction (N-S) in which the source shows double
structure in ISO maps (see later).
   
\begin{figure*}
\centering
\includegraphics[width=16.0cm]{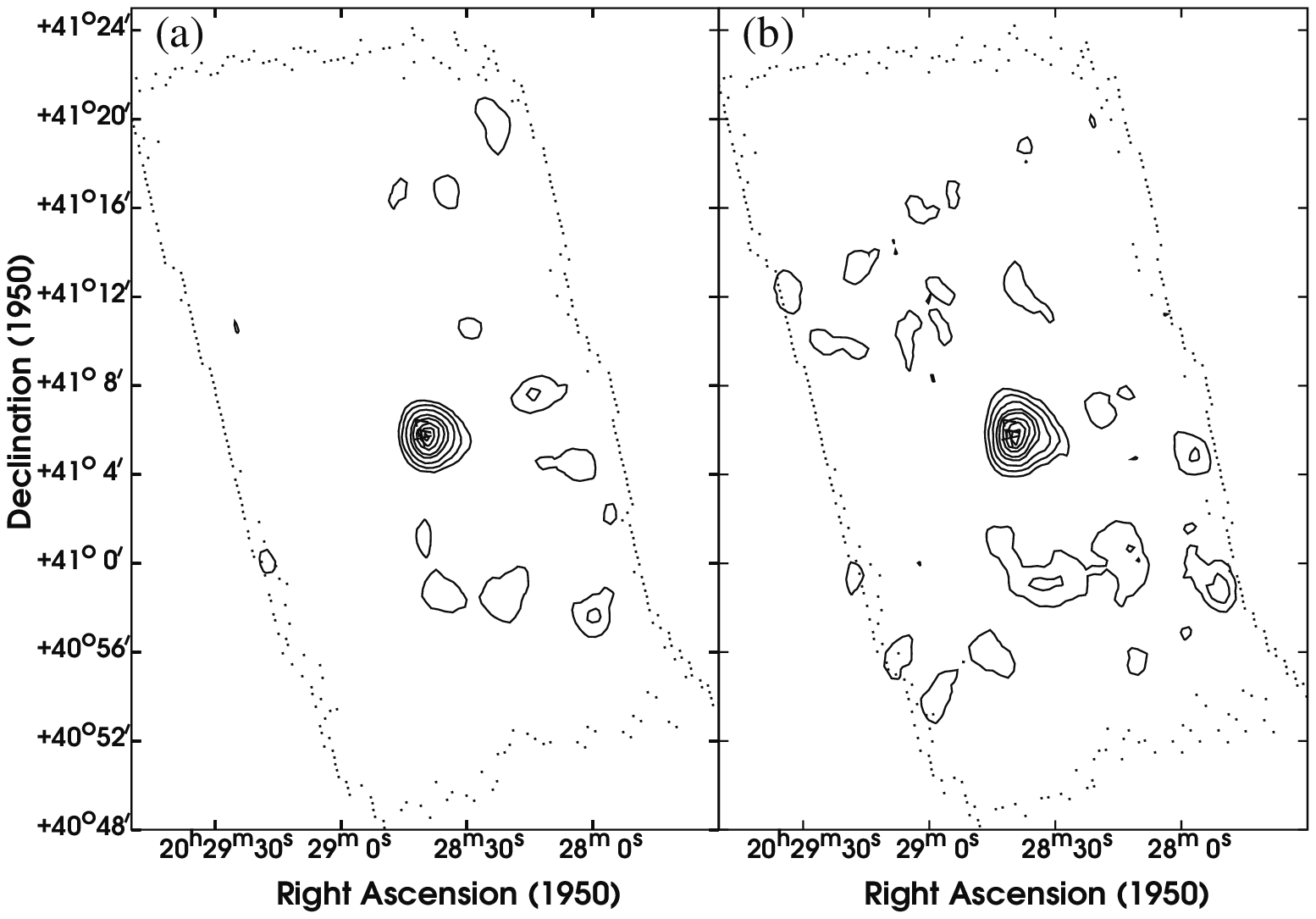}
   \caption{Same as figure 1 but for the region around IRAS 20286+4105.
The peak intensities are 770 Jy arcmin$^{-2}$ at 150 $\mu$m and 402 Jy
arcmin$^{-2}$ at 210 $\mu$m.}

\end{figure*}

 The dust temperature (T$_{d}$) and optical depth ($\tau_{150}$) maps for 
IRAS 20286+4105 are shown in Fig. 10. The dust in IRAS 20286+4105 is very
much cooler (temperature range of 15 K to 25 K) than the other two sources.
This is perhaps because the luminosity for this source is lowest (see Table
4). The peak in temperature occurs near the peak of the flux density
distribution.  However the peak in optical depth is shifted  to the
north by $\sim 1 \arcmin$.

\begin{figure*}
\centering
\includegraphics[width=16.0cm]{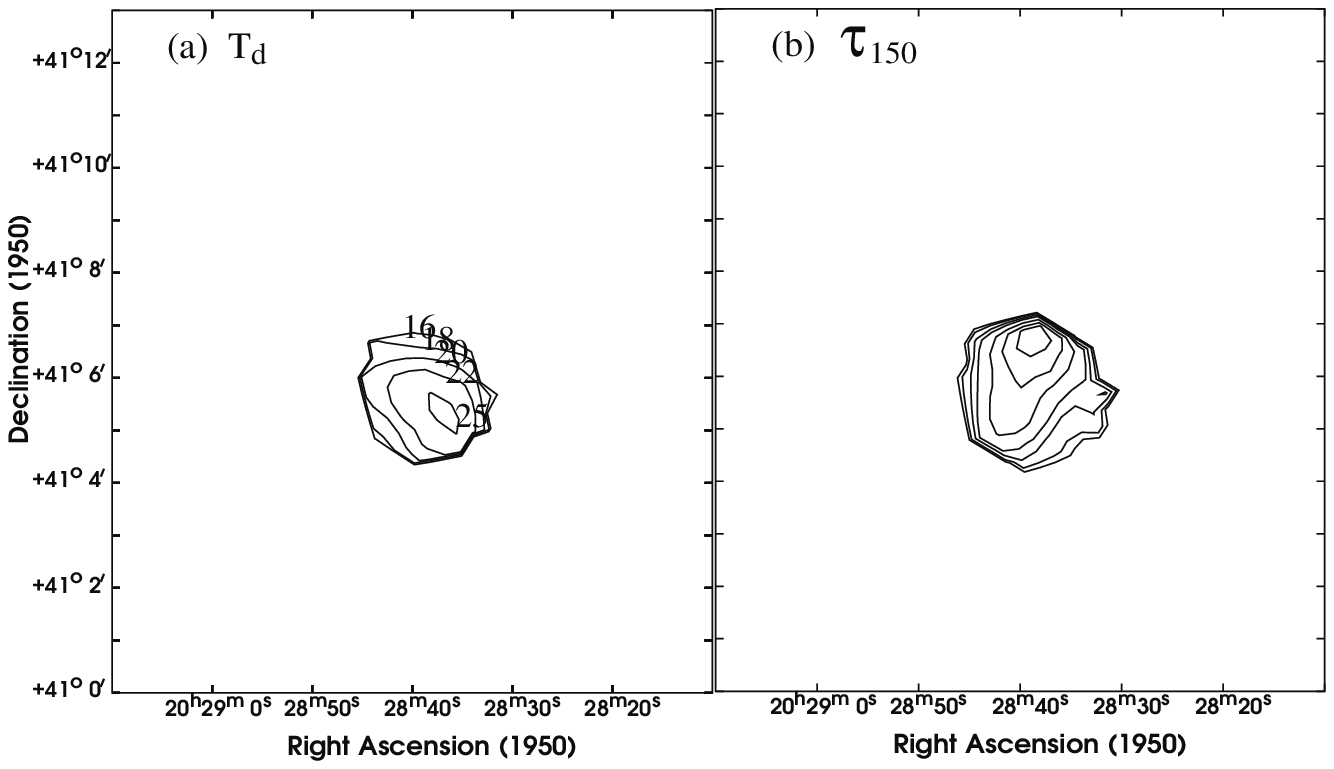}

\caption{Same as figure 2 but for the region around IRAS 20286+4105.
The temperature contours are at the levels of 25, 22, 20, 18 and 16 K
and for the optical depth distribution the peak value is 0.042.}

\end{figure*}

   The HIRES processed maps of IRAS 20286+4105 in the four IRAS bands,
are shown in Fig. 11. As seen from Table 1, the source is only resolved at
12 $\mu$m and partly resolved at 60 $\mu$m.

\begin{figure*}
\centering
\includegraphics[width=16.0cm]{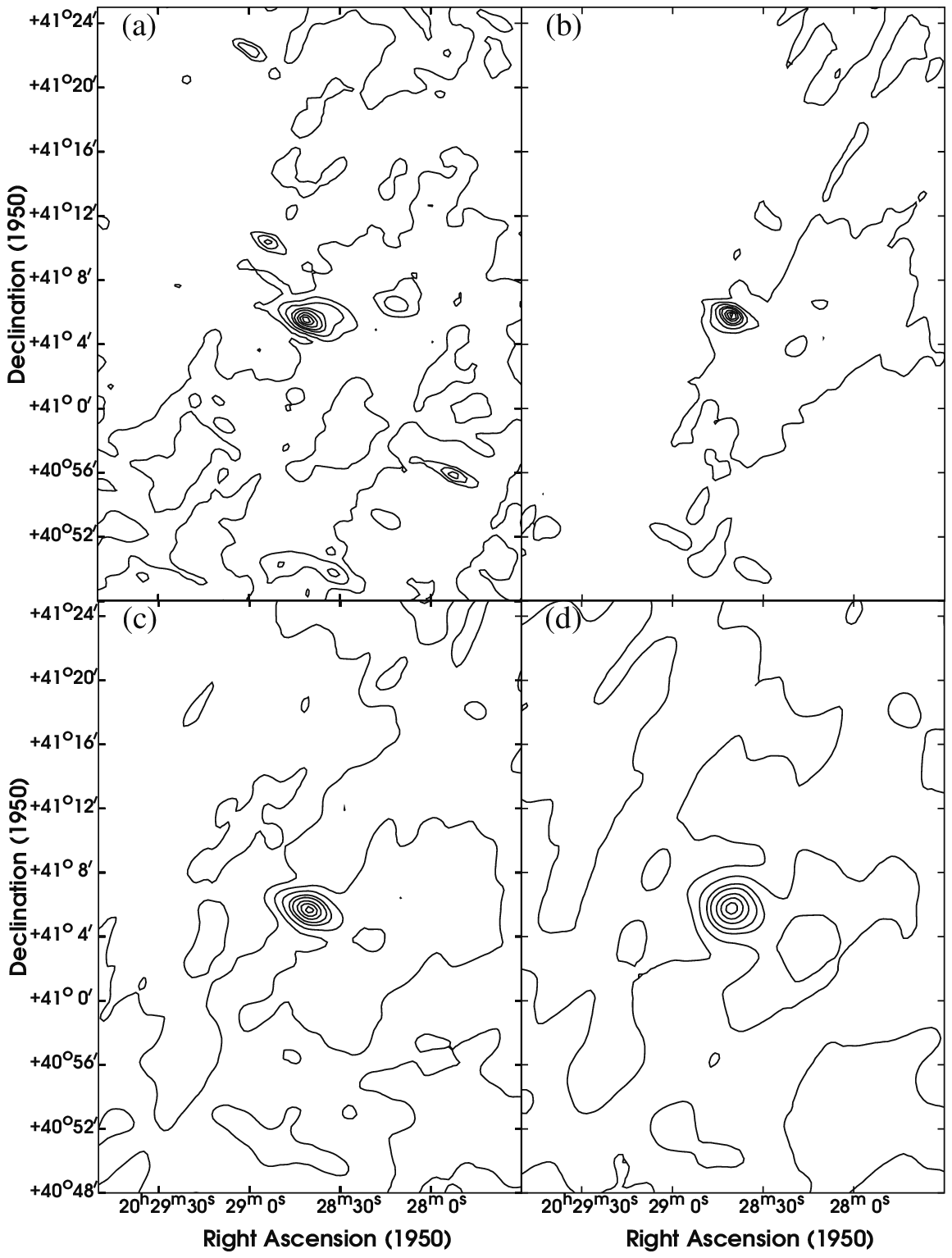}

   \caption{The HIRES processed intensity maps of the region around IRAS
20286+4105 at {\bf (a)} 12 $\mu$m, {\bf (b)} 25 $\mu$m, {\bf (c)} 60 $\mu$m
and {\bf (d)} 100 $\mu$m. 
The contour levels are at  0.90, 0.70, 0.50, 0.30, 0.20, 0.10, 0.05 
and 0.025 of the peak intensity. The peak intensities are 22.1, 175, 509
and 362 Jy arcmin$^{-2}$ at 12, 25, 60 and 100 $\mu$m respectively.}

\end{figure*}

    Maps generated from the ISOCAM images in the seven bands are shown in
Fig. 12. It can be seen from this figure that IRAS 20286+4105 consists of
two main components, separated along the N--S direction by $\sim32\arcsec$
in all the filters. There is a third weaker component to the east. The
structure in our maps (especially the SW maps) is very similar to the
structure in the K band image of Comeron and Torra (2001).

\begin{figure*}
\centering
\includegraphics[width=17.0cm]{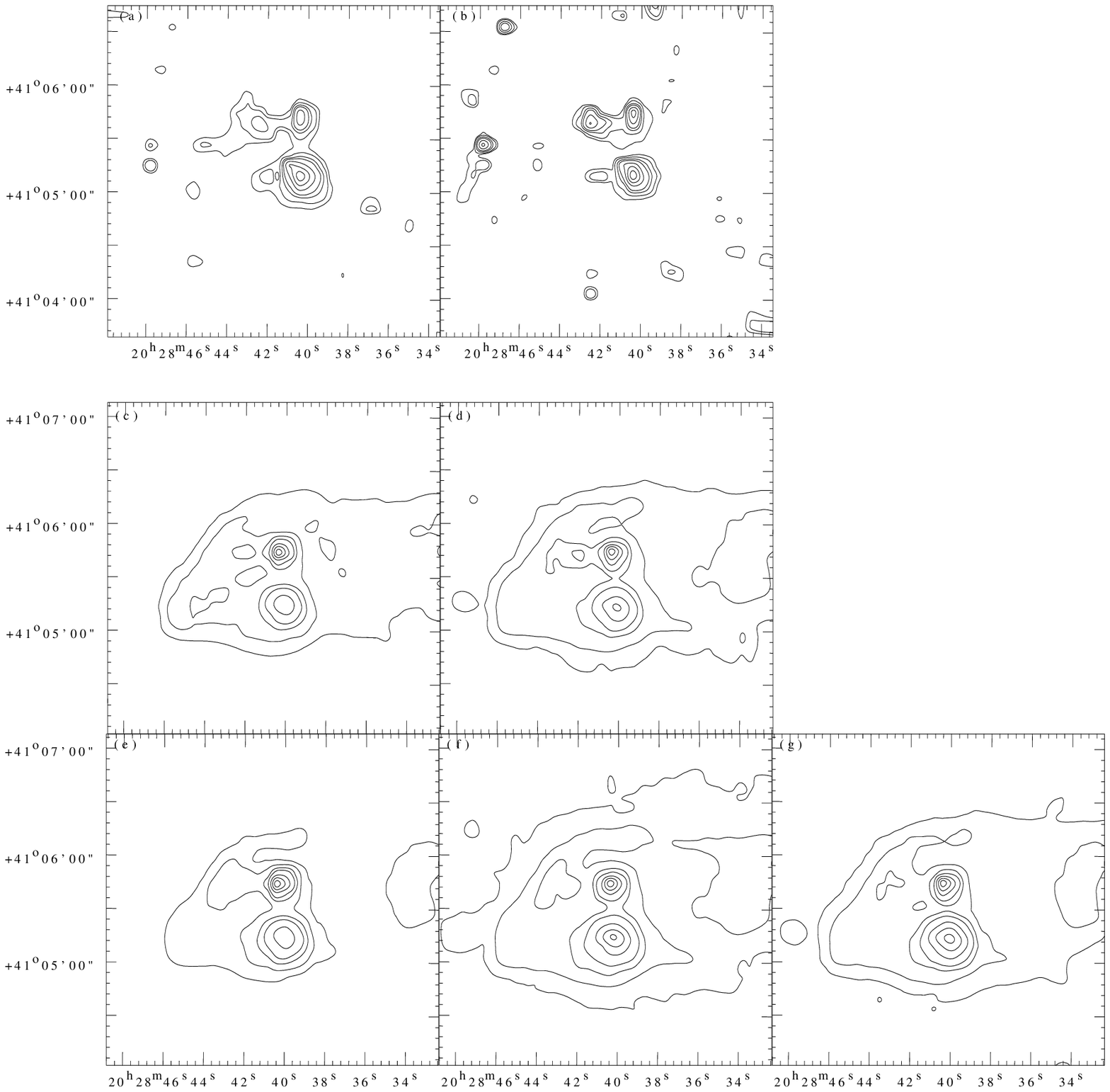}

\caption{Same as figure 4 but for the region around IRAS 20286+4105. 
The peak intensities are 9.9, 4.4, 77.4, 70.7, 183, 41.7 and 68.2 Jy
arcmin$^{-2}$ at 3.30, 3.72, 6.00, 6.75, 7.75, 9.62 and 11.4 $\mu$m
respectively.}
\end{figure*}

\section{Radiation Transfer Calculations}

  To understand the infrared emission from these sources radiation transfer
calculations have been performed. The details of the basic radiation transfer
scheme are given in Mookerjea et al. (1999). The geometry of the basic model
is shown in Fig. 13a. Briefly, an internal radiation source, i.e., a star is
surrounded by a spherically symmetric envelope of dust and gas mixture.
The gas is assumed to be hydrogen and surrounds the star from the stellar
surface itself. The dust is a mixture of astronomical silicates
and graphite and there is an inner cavity of radius R$_{min}$ which is
devoid of dust. Two types of interstellar dust are considered; those with
absorption and emission properties given by Draine \& Lee (1984) (hereafter
DL) and those with properties given by  Mathis et al. (1983) (hereafter
MMP) and the size distribution of the grains is taken to be a power law
given by Mathis et al. (1977), viz. n(a)da $\propto$ a$^{-m}$da, with m
= 3.5 and 0.01 $\mu$m $\leq$ a $\leq$ 0.25 $\mu$m; a is the radius of the
grain. The envelope is surrounded by the interstellar radiation field (ISRF)
taken from Mathis et al. (1983). The radial density distribution
of the gas and dust has been taken to be a power law, i.e. n(r) $\propto$
r$^{-\alpha}$ with $\alpha$ = 0, 1 or 2. The radiation transport has been
carried out using a program based on the code CSDUST3 (Egan et al. 
1988). The parameters of the model are -- inner radius of the dust
cloud R$_{min}$, outer radius of the envelope R$_{max}$, $\alpha$,
luminosity and temperature of the embedded star, relative abundance of the
two types of dust, gas to dust ratio and the total radial optical depth at
a selected wavelength ($\tau_{100}$ at 100 $\mu$m). The radiation transfer
of Lyman continuum photons in gas is carried out in a consistent manner
with that through the dust. The combination of the parameters is obtained
which gives lowest $\chi^{2}$ to fit the observed SED. The scheme also gives
the expected radio emission and the radial distribution of infrared emission.
These have also been used to constraint the model. This basic scheme was
used to model IRAS 20178+4046. 

  As ISOCAM maps of IRAS 20286+4105 show two main components in all the
images, the radiation transfer calculations for this source were performed
differently. Whereas the basic scheme of the radiation transfer is the same,
it is assumed that there are two cores. Each core, with its own internal
radiation source, is surrounded by a spherically symmetric envelope of dust
and gas mixture. The type of dust (DL or MMP) and the silicate to
graphite ratio is assumed to be the same for both cores. Total
luminosity of the source is divided between the two cores in the ratio
of their respective ISOCAM flux densities. The radiation transfer calculation
is done for each core separately and the radiation emerging from each core
is calculated. These cores are surrounded by another common shell, with the
internal input radiation equal to the sum of the emerging radiation from each
core. This common shell is surrounded by interstellar radiation field taken
from Mathis et al. (1983). The radiation emerging from the inner boundary of
this shell is the input at the outer boundary of the cores. The calculations
are done iteratively. The geometry of the model is shown in Fig. 13b.

   The radiation transfer calculations for IRAS 19181+1349 were performed
in a cylindrical geometry with two embedded sources. This model has been
described in Karnik and Ghosh (1999). An attempt was made to fit
the SED as well as the radial profile of the intensities.

\section{Discussion}

\subsection{IRAS 20178+4046}

First we consider IRAS 20178+4046 which is the simplest of the three
sources because it shows only one component in all the wavelength bands. 
We have constructed the SED of IRAS 20178+4046 using the flux densities from
balloon-borne, IRAS and ISO observations, given in Tables 2 and 3, IRAS LRS
data from Volks and Cohen (1989) and ground based observations of Faison et
al. (1998). This SED is shown in Fig. 14a. In this figure, we have only
shown ISOCAM flux densities in the continuum bands; PAH bands are treated
separately. It can be seen that while there is general agreement between
various observations, HIRES flux density at 12 $\mu$m is higher than IRAS
LRS data as well as the data of Faison et al. (1998). This is because the
HIRES flux density is integrated over 4$\arcmin$ dia and includes substantial
amount of extended flux. Also, the flux densities of Faison et al. (1998)
are lower than those of IRAS LRS as well as from ISO observations. This is
because of the small aperture (9$\arcsec$) of Faison et al. (1998).

%
\begin{figure*}
\centering
\includegraphics[width=17.0cm]{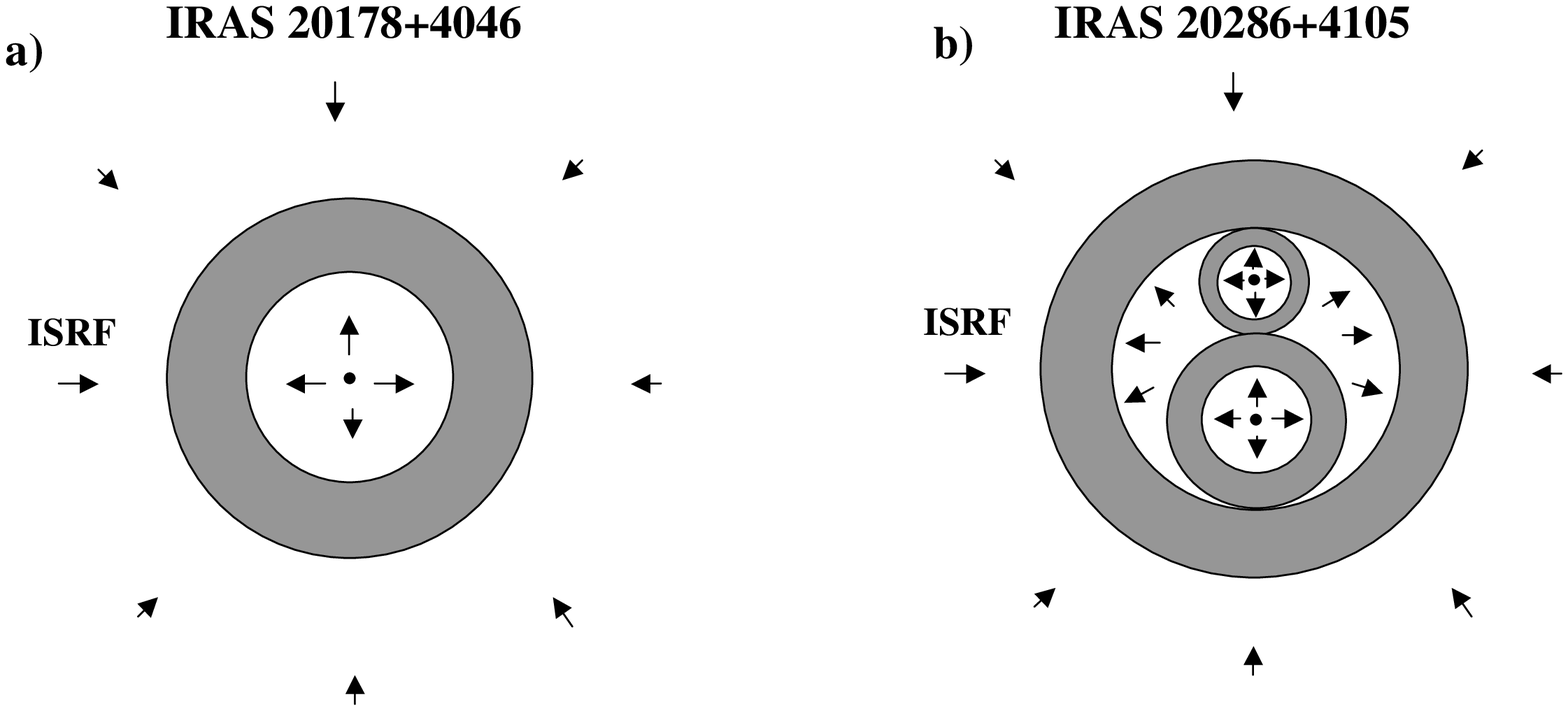}
\caption{ Geometry of the models for radiation transfer calculations;
{\bf (a)} for IRAS 20178+4046, {\bf (b)} for IRAS 20286+4105. The dark
circles in the centre represent embedded central stars and the grey
areas the spherical shells of dust and gas. The white areas between the
two contain only gas. For IRAS 20178+4046, the shell is heated from outside
by the ISRF and from inside by the radiation from the embedded star. For IRAS
20286+4105, the radiation emerging out of the two inner gas-dust shells is
combined and then used for heating the outer shell from inside. For more
details refer to the text.
}
\end{figure*}
%
\begin{figure}
\includegraphics[width=6.3cm]{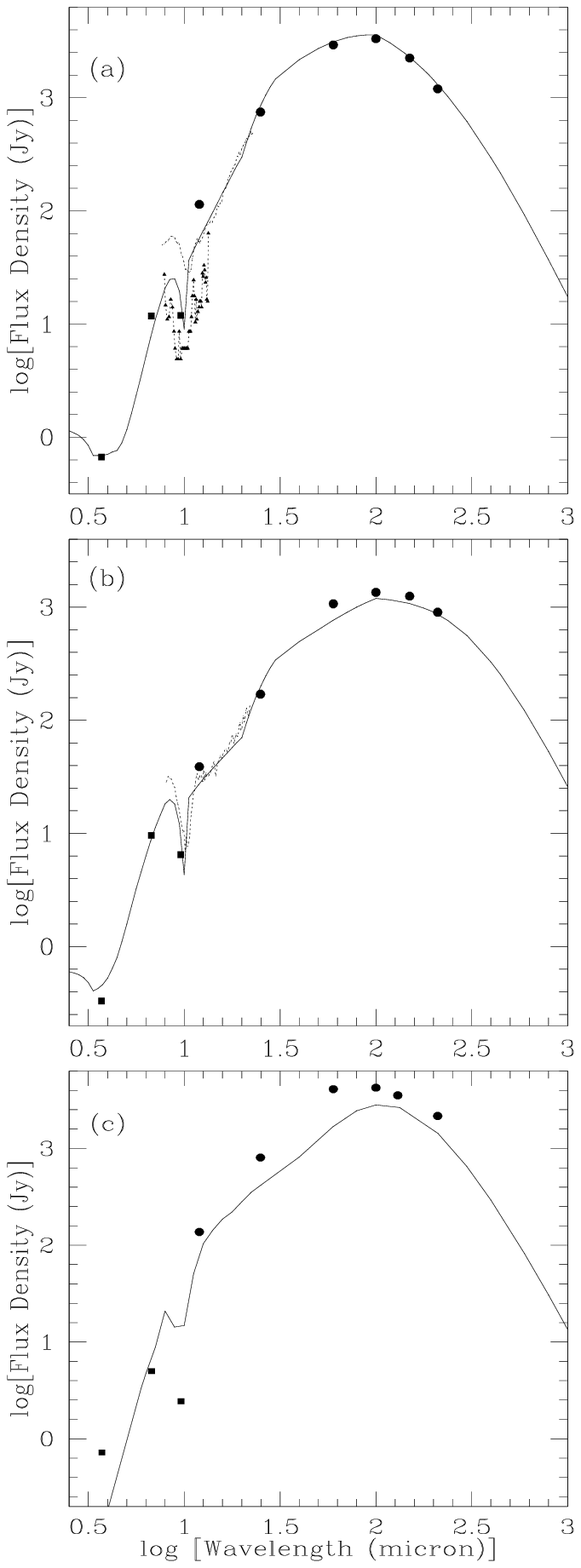}
\caption{ {\bf (a)} SED of IRAS 20178+4046. Symbols are -- circles -
flux densities from balloon-borne observations and IRAS HIRES,
squares - from ISOCAM maps, triangles - from ground based observations of
Faison et al. (1998) and dotted line - from IRAS LRS. The solid curve
shows the SED corresponding to the best fit model. {\bf (b)} Same as
{\bf (a)} but for IRAS 20286+4105. {\bf (c)} Same as {\bf (a)} but for IRAS
19181+1349. The solid curve shows the best fit model for this source given
in Karnik and Ghosh (1999).
}
\end{figure}

The SED from the best fit radiation transfer model is shown in Fig. 14a.
It can be seen that the calculated SED fits the observed one quite well.
The parameters of the best fit model are given in Table 4. The luminosity
corresponds to an O7 ZAMS star. The density distribution that fits best is,
n(r) $\propto$ r$^{-2}$, the dust type is MMP and is dominated by silicates.
The calculated angular sizes (deconvolved with the angular resolutions) are
given in Table 1. It can be seen that the calculated angular sizes are
consistent with the observed sizes. The calculated radio flux density at
6 cm is 72 mJy for a gas to dust ratio of 100. This can be compared with
the observed flux density of 69 mJy by McCutcheon et al. (1991) and 65 mJy
by Wilking et al. (1989). Density distribution (r$^{-2}$) is consistent
with what is expected in the envelopes of young protostars from
free fall models (see for example Shu et al. 1987). While it is true that 
the fitting of SED alone is not unique (Churchwell et al. 1990), in the
present case angular distribution and radio flux are consistently modelled
and therefore parameters of the model are on firmer footing.

\subsection{IRAS 20286+4105}

We have constructed the SED for IRAS 20286+4105 using the flux densities
from balloon-borne, IRAS and ISO observations, given in Tables 2 and 3
and IRAS LRS data from Olnon et al. (1986). This SED is shown in Fig. 14b.
For ISO observations we have combined the flux densities of two
components (N and S); the third component toward the East gets weaker at
longer wavelengths and therefore has not been included. It can be seen
that there is good agreement between various observations.

The SED from the best fit radiation transfer model is shown in Fig. 14b. 
The calculated SED fits the observed SED quite well. The parameters of the
best fit model are given in Table 4. Here the two cores are denoted as IRAS
20286+4105 N and IRAS 20286+4105 S. The luminosities of the two cores
correspond to ZAMS stars of the type B0.5 (N) and B0-0.5 (S). The dust
density distribution that fits best is, n(r) $\propto$ r$^{-1}$ and the
dust type is again MMP and dominated by silicates. The radiation transfer
model predicts negligible radio emission. No compact radio source has been
observed in the field of IRAS 20286+4105. McCutcheon et al. (1991) see an
extended feature at 6 cm and according to them the association of radio
continuum with IRAS 20286+4105 is not clear. Similarly Wendker et al. (1991)
see radio emission from IRAS 20286+4105 as a small density enhancement on a
large ridge. Thus the model is consistent with radio observations. The
flatter dust density distribution, as compared to IRAS 20178+4046, suggests
that IRAS 20286+4105 is relatively more evolved system.

\begin{table*}
\caption{Parameters$^*$ for the best fit radiation transfer models}
\begin{center}
\begin{tabular}{llllllll}
\hline
Source & R$_{max}$  & R$_{min}$  & L & $\alpha$ & $\tau_{100}$ &
Dust Type & f$_{Silicate}$ \\
 & (pc) &  (pc)  & (10$^{4}$~L$_{\sun}$) & & & & \\
\hline

 20178+4046 & 3.40 & 0.077 & 9.5 &  2 & 0.12 &   MMP  & 0.76  \\
 & & & &  &  & & \\

 20286+4105 N & 0.24 & 0.022  & 1.2 & 1 & 0.11 &   MMP  & 0.65    \\

 20286+4105 S & 0.32 & 0.022 & 2.3 &  1 & 0.11 &   MMP  & 0.65    \\

 20286+4105 Outer & 3.02 & 1.12 & 3.5 & 1 & 0.025 &   MMP  & 0.65  \\
 & & & &   &  & & \\

 19181+1349  & 1.8 & 0.03, 0.01 & 62, 63& 0 & 0.10 &   DL  & 0.80    \\
\hline
\end{tabular}
\end{center}
  $^*${\footnotesize The scheme of calculation for IRAS 19181+1349 is
different from that of other sources. Therefore some of the parameters
are not directly comparable.}\\

\end{table*}

\subsection{IRAS 19181+1349}

The SED for IRAS 19181+1349 using the flux densities from balloon-borne,
IRAS and ISO observations, is shown in Fig. 14c. As was mentioned earlier,
maps of IRAS  19181+1349 show multiple components in all the images.
For this source we have used the flux density within  5$\arcmin$ of the main
peak such that the contribution from multiple sources is included. 

   The SED for IRAS 19181+1349 was analyzed by Karnik and Ghosh (1999)
using a model with two embedded sources in a cylindrical geometry.
An attempt was made to fit the SED as well as radial profile of the
intensities. The best fit to the SED is shown in Fig. 14c. The parameters
of that model are given in Table 4. The prefered dust density distribution
(r$^0$) suggests that IRAS 19181+1349 is very evolved system.

\subsection{PAH emission}

       As seen from Table 3, the emission in PAH bands is generally higher
as compared to the neighbouring bands. We have derived line fluxes in the
PAH bands from the flux densities given in Table 3. For this, we have
subtracted the continuum flux density taken from the neighbouring filter;
SW6 (3.72 $\mu$m) for 3.30 $\mu$m band, LW5 (6.75 $\mu$m) for 6.2 and 7.7
$\mu$m bands and LW7 (9.6 $\mu$m) for 11.3 $\mu$m band; and mutiplied by the
filter bandwidth. This procedure will lead to an overestimation for the
11.3 $\mu$m PAH feature because the continuum given by LW7 is underestimated
due to silicate absorption. The band fluxes and the band ratios are
given in Table 5. It is seen that the strongest PAH feature is the one at
7.7 $\mu$m whereas the feature at 6.2 $\mu$m is absent in most of the cases.
We have also given the range of band ratios obtained by Roelfsema et al.
(1996) using ISO Short Wavelength Spectrometer (SWS) for six compact
H\,{\sc ii} and the average value of line ratios from typical H\,{\sc ii}
regions from Cohen et al. (1989). It is seen that the band ratios obtained
by us are consistent with the values of Roelfsema et al. (1996) for three of
the four features observed by us. However they are different from the values
of Cohen et al. (1989) for typical H\,{\sc ii} regions. For example, the
ratio of 7.7 $\mu$m and 3.3 $\mu$m features varies from 18 to 52 in our data
as compared to 17 to 73 in those of Roelfsema et al. (1996), but the average
value of Cohen et al. (1989) is 7. This shows that there is a difference in
the PAH emission by compact H\,{\sc ii} regions and other H\,{\sc ii}
regions. However, we can not understand the near absence of the 6.2 $\mu$m
PAH band in our observations especially as both, the 6.2 $\mu$m feature and
the strongest observed feature at 7.7 $\mu$m arise from C-C stretching and
have been found to be strongly correlated (Cohen et al. 1989).

\begin{table*}[tb]
  \caption{PAH band fluxes for the sources in ISOCAM images}
  \begin{center}
    \leavevmode
  {\footnotesize
\begin{tabular}{lccccccccc} \hline
IRAS& RA & Dec. & \multicolumn{4} {c}{Band Flux$^{a}$ (10$^{-11}$ 
erg cm$^{-2}$ s$^{-1}$)} & \multicolumn{3} {c}{PAH Band Ratios}\\
\cline{4-10}
Source & (1950) & (1950) & 3.3\,$\mu$m & 6.2\,$\mu$m & 7.7\,$\mu$m &
  11.3\,$\mu$m & 6.2/3.3 & 7.7/3.3 & 11.3/3.3\\
     \hline
19181+1349  & 19 18 05.4  &+13 49 22 & -- & --   & 9.8   & --    &   --
& -- & --\\
            & 19 18 06.8  &+13 49 40 & -- & --   & 9.1   & 5.2   &  -- &
 -- & -- \\
            & 19 18 11.2  &+13 49 33 & -- & --   & 14.4  & --    & --  &
 --  &  -- \\ 
            & 19 18 12.5  &+13 51 24 & 0.06 & -- & --    & --    & --  &
--  & --  \\ 
            & 19 18 12.6 & +13 50 10 & 1.16 & -- & 23    & 4.6   & -- &
20  & 3.9\\ 
           &19 18 13.0  &+13 49 59 & 0.55 & --   & 10.1  & --    & -- &
 18 & -- \\
            & 19 18 13.5 & +13 49 24 & -- & --   & 7.9   & --    & -- &
-- & --\\ 
  &  &  & & & & & & & \\ 
20178+4046 & 20 17 53.3 & +40 47 03 &  6.1 & 7.6 & 149   & 30    & 1.25 &
25 & 5.0 \\       
          & 20 17 55.0 & +40 47 01  &  -- & --   & 36    & 5.8   & -- &
 -- & --\\
  &  &  & & & & & & & \\ 
20286+4105 & 20 28 40.3 & +41 05 14 & 2.6 & --   & 51    & 8.5   & -- &
19 & 3.2 \\ 
           & 20 28 40.3 & +41 05 45 & 0.66 & --  & 34    & 2.6   & -- &
52 & 4.0\\ 
           & 20 28 42.5 & +41 05 44 & 0.44 & 4.2 & 8.3   & 0.91  & 9.5 &
19 & 2.1\\ 
&  &  &  & & & & & & \\ 
 
Compact H\,{\sc ii}  & & & & & & & 3.4 -- 22 & 7 -- 44 &
2.9 -- 6.7\\
regions$^{b}$ & &  &  & & & & & & \\ 
&  &  &  & & & & & & \\ 

Typical H\,{\sc ii}& &  & & & & & 4.0 & 7 &  2.3\\
regions$^{c}$  & &   & & & & & & & \\ 
\hline
\end{tabular}
}
\newline
  \end{center}
  $^a${\footnotesize Band fluxes have been caculated after subtracting the
continuum defined by the neighbouring filter as given in the text}\\
$^b${\footnotesize The flux ratios for compact H\,{\sc ii} regions from
Roelfsema et al. (1996)}\\
$^c${\footnotesize The flux ratios for typical H\,{\sc ii} regions from
Cohen et al. (1989)}\\

\end{table*}

In addition, an attempt has been made to generate the spatial distribution
of emission in individual PAH features for the regions imaged by ISOCAM
filters. The PAH emission maps have been generated by subtracting out the
expected continuum emission. The spectrally local continuum emission has
been estimated by linear interpolation or extrapolation using the
neighbouring ISOCAM filters (viz., LW5 \& LW7 for the features at 7.7
$\mu$m and 11.3 $\mu$m features; LW5 \& SW6 for the features at 3.3 $\mu$m
and 6.2 $\mu$m). Generally the above method has been successful, except for
the case of 3.3 $\mu$m feature for IRAS 20178+4046 and IRAS 20286+4105. The
failure of the latter cases is attributed to a large ratio ($\sim$ 15)
of intensities between 6.75 $\mu$m and 3.72 $\mu$m, making linear
extrapolation a suspect. In these cases the measured intensities 
in the nearest filter (SW6) itself have been used as an estimate of
the local continuum.

 The spatial distribution of emission in individual PAH features is
found to be diffuse and widespread in nature. However, they qualitatively
(morphologically) resemble the isophotes of total emission in the
corresponding ISOCAM filters (i.e. uncorrected for the continuum),
which have been  presented in figures 4, 8 and 12. Hence these are
not presented here. Instead, some quantitative results are presented.
The total emission in the individual PAH features from the imaged
regions of IRAS 19181+1349, 20178+4046 and 20286+4105 are tabulated in
Table 6.

\begin{table*}
\begin{center}
\caption{Total emission in individual PAH features extracted from ISOCAM
data}
\vskip 0.5cm
\begin{tabular}{cccc}
\hline
PAH & \multicolumn{3}{c}{Total emission in the PAH feature}\\
feature & \multicolumn{3}{c}{ erg sec$^{-1}$ cm$^{-2}$ } \\
\cline{2-4}
& IRAS 19181+1349& IRAS 20178+4046 & IRAS 20286+4105 \\

\hline

 3.3 $\mu$m & 
 1.56 $\times 10^{-9}$& 1.15 $\times 10^{-10}$ & 5.20 $\times 10^{-11}$ \\

 6.2 $\mu$m & 
  -- & 3.44 $\times 10^{-10}$ & -- \\

 7.7 $\mu$m & 
 6.03 $\times 10^{-9}$& 4.01 $\times 10^{-9}$ & 1.56 $\times 10^{-9}$ \\

 11.3 $\mu$m & 
 7.45 $\times 10^{-10}$& 9.62 $\times 10^{-10}$ & 3.60 $\times 10^{-10}$ \\

\hline
\end{tabular}
\end{center}
\end{table*}

 Considering that the ratios of emission in the individual PAH features
 may contain information about their excitation mechanisms, these have been
 computed on pixel-by-pixel basis for the three sources considered here. 
 The determination of these ratios have been restricted to regions with
 sufficient signal strengths in relevant PAH features. The summary of
 these ratios are presented in Table 7.

  The scatters in the values of the feature ratios (see the
 standard deviations) is not measurement noise but represent 
 their genuine spatial variation. This implies different physical
 conditions around each of the star forming regions which 
 differentially affect
 the exciting mechanisms of individual PAH features.
 Whereas the mean ratios are similar for 
 the regions  IRAS 20178+4046 and 20286+4105, the values for
 the region  IRAS 19181+1349 differ significantly.
 This may be related to their evolutionary stages, which is
 also indicated by the conclusions from the radiative
 transfer modelling (see Sect. 4.1.3; viz., the power law index
 of radial density distribution is flattest for IRAS 19181+1349,
 indicating it to be the most evolved source).

\begin{table}
\begin{center}
\caption{ Distribution of PAH feature ratios for the ISO images}
\vskip 0.5cm
\begin{tabular}{cccc}
\hline
IRAS & $I_{11.3}$/$I_{7.7}$ & $I_{7.7}$/$I_{6.2}$ & $I_{7.7}$/$I_{3.3}$ \\
Source\\

\hline
 19181+1349 &  0.30$\pm$0.39$^{*}$ &     -- &  23.2$\pm$11.8$^{*}$ \\
            &   (1024)$^{**}$    &        &          (476)$^{**}$ \\
 
 20178+4046 &  0.84$\pm$0.40 &   14.7$\pm$8.3 &    19.9$\pm$14.0\\
                 &   (289)     &     (199)   &     (143) \\

 20286+4105 &  1.07$\pm$0.64  &    --     &     13.4$\pm$11.7\\ 
                 &   (256)      &            &     (145) \\

\hline
\end{tabular}
\end{center}

$^{*}${\footnotesize  mean $\pm$ std. dev.}\\
$^{**}${\footnotesize  (sample size : \# of pixels)}\\

\end{table}

\section{Conclusions}

We have mapped two ultracompact H\,{\sc ii} regions and one molecular clump 
simultaneously in two far infrared bands using 1 m TIFR balloon-borne 
telescope. Using the maps from the balloon-borne observations, maps of dust
temperature and optical depth have been obtained. We have also imaged the
central $3\arcmin \times 3 \arcmin$ regions of these sources in seven mid
infrared bands using ISOCAM instrument of ISO. IRAS HIRES processed data
have also been obtained for these sources in the four IRAS bands. There are
certain similarities between various maps of the sources. Using the flux
densities in all the 13 bands, as well as any other available data, SEDs
have been constructed. Radiation transfer calculations have been done for
the three sources with different geometries.

All the maps of IRAS 20178+4046 consist of mainly one source, with a lobe
seen in the ISOCAM maps. This source has been modelled with a single core
and excellent fit has been obtained to the SED, radio flux and angular sizes.
The dust density distribution which gives best fit to the data is of the
form r$^{-2}$ and the dust is dominated by silicates.

While the IRAS HIRES and TIFR maps of IRAS 20286+4105 consist of one
source, ISOCAM maps show three sources at shorter wavelengths and two at
longer wavelengths. This source has been modelled with two cores
and excellent fit has been obtained to the SED along with consistency with
radio observations. Dust in IRAS 20286+4105 has been found to be coolest.
The dust density distribution which gives best fit to the data is of the
form r$^{-1}$ and the dust is dominated by silicates.

The maps of IRAS 19181+1349 show multiple sources in all wavelengths.
Hot spots have been found at positions away from the intensity peaks.
Radiation transfer calculations for this source have been done in
cylindrical geometry with two cores. The dust distribution which
gives best fit to the data is the one with uniform density i.e. r$^{0}$ and
the dust is dominated by silicates.

 The differing density distributions tend to imply that IRAS 20178+4046, 
20286+4105 and 19181+1349 are in progressive order of their evolutionary
stage.

Fluxes in four PAH bands have been obtained from the ISOCAM images. Ratios
of the emission in different PAH bands have been obtained. Whereas two out
of the three ratios have been found to be similar to the ratios obtained for
other compact H\,{\sc ii} regions the third ratio is quite different. Also,
the band ratios for compact H\,{\sc ii} regions have been found to be
different from those for general H\,{\sc ii} regions. Spatial
distribution of PAH emission has been obtained and found to be diffuse
and widespread.

\begin{acknowledgements}  
  We thank C.B. Bakalkar, S.L. D'Costa, S.V. Gollapudi, G.S. Meshram, M.V.
Naik, M.B. Naik and D.M. Patkar for valuable technical support for the
program. We thank the members of the Balloon Support Instrumentation Group
(TIFR) and the National Balloon Facility, Hyderabad (TIFR) for providing
support for the balloon flight. We thank the Infrared Processing and Analysis
Center, Caltech for providing the HIRES processed IRAS data. We thank the ISO
Observing Time Allocation Committee for alloting time for observations with
ISO.

\end{acknowledgements}

\end{document}